# Unusual effects of anisotropy on the specific heat of ceramic and single crystal MgB$_2$

F. Bouquet†, Y. Wang†, I. Sheikin†, P. Toulemonde†,

M. Eisterer‡, H. W. Weber‡, S. Lee§, S. Tajima§, and A. Junod†

† DPMC, University of Geneva, CH-1211 Geneva 4, Switzerland
‡ Atomic Institute of the Austrian Universities, A-1020 Vienna, Austria
§ Superconductivity Research Laboratory, ISTEC, Tokyo 135-0062, Japan

## Abstract

The two-gap structure in the superconducting state of MgB$_2$ gives rise to unusual thermodynamic properties which depart markedly from the isotropic single-band BCS model, both in their temperature- and field dependence. We report and discuss measurements of the specific heat up to 16 T on ceramic, and up to 14 T on single crystal samples, which demonstrate these effects in the bulk. The behavior in zero field is described in terms of two characteristic temperatures, a crossover temperature $T_{c,\pi} \cong 13$ K, and a critical temperature $T_c = T_{c,\sigma} \cong 38$ K, whereas the mixed-state specific heat requires three characteristic fields, an isotropic crossover field $\mu_0 H_{c2,\pi} \cong 0.35$ T, and an anisotropic upper critical field with extreme values $\mu_0 H_{c2,\sigma,c} \cong 3.5$ T and $\mu_0 H_{c2,\sigma,ab} \cong 19$ T, where the indexes $\pi$ and $\sigma$ refer to the 3D and 2D sheets of the Fermi surface. Irradiation-induced interband scattering tends to move the gaps toward a common value, and increases the upper critical field up to ~ 28 T when $T_c \cong 30$ K.

**Corresponding author**

Alain Junod
Département de physique de la matière condensée
24, quai Ernest-Ansermet
CH-1211 Geneva 4
(Switzerland)
phone: +41 22 702 6204
fax:   +41 22 702 6869
e-mail: alain.junod@physics.unige.ch

## I. Introduction

Shortly after the unexpected discovery of superconductivity at ~ 39 K in MgB$_2$ [1, 2], specific heat (*C*) measurements were performed on ceramics in order to learn more about the nature of superconductivity in this material [3-10]. After 15 years of investigations of high temperature superconductors (HTS), superconductivity above ~ 30 K was intuitively associated with unconventional pairing, possibly mediated by antiferromagnetic fluctuations, involving short coherence lengths ξ, reduced dimensionality, and thermal fluctuations. In conventional superconductors, specific heat experiments, with their ability to respond globally to those characteristic excitations with energies of the order of $k_BT$, have provided from the very beginning strong hints and support to the classic theory of superconductivity [11]. They have later given important information on specific features of HTS, such as critical fluctuations, melting of the vortex lattice, presence of line nodes in the gap, etc. (see e.g. Refs. [12, 13] and references therein). Information obtained from *C* lacks the resolution of modern spectroscopic tools such as scanning tunneling spectroscopy (STS) or angle-resolved photoemission spectroscopy (ARPES), but, in conjunction with the latter, can establish that unconventional features are not limited to the surface of the sample, but are bulk properties.

The basic characterization of MgB$_2$ brought about both disappointment and surprises. Measurements of the isotope effect established that the interaction responsible for the formation of pairs is mediated by phonons [4, 14], and nuclear magnetic resonance showed that the symmetry of Cooper pairs is s-wave [15]. Therefore, MgB$_2$ appeared as a somewhat "uninteresting" superconductor. Determinations of the phonon density of states [16], and the analysis of the lattice specific heat up to room temperature [9], showed that average phonon frequencies were typically 2-3 times as high as for classic superconductors, such as Nb$_3$Sn (Table I). This could have possibly accounted for the higher critical temperature, if the electron-phonon coupling strength as measured by the parameter $\lambda_{e-ph}$ had been of the same order – but it was not. This was conclusively demonstrated by band-structure calculations of the density of states (DOS) at the Fermi level [17-22], compared to measurements of the renormalized DOS as measured by the Sommerfeld coefficient $\gamma_n$ of the specific heat [3, 6, 9, 10]. The mass renormalization factor, $1+\lambda_{e-ph}$, is small (Table I), leaving the origin of the high critical temperature unexplained within the standard *i*sotropic *s*ingle-*b*and BCS (ISB) theory. The relatively small values of $\gamma_n$ and of the condensation energy $E_c \equiv \mu_0 H_c^2/2$ determined by specific heat experiments (Table I) also appeared to be inconsistent with superconductivity near ~ 40 K.

As it is now understood, most of these puzzles are associated with an improper use of average quantities in the analysis of a highly anisotropic, but otherwise classic superconductor. By anisotropy we mean that both the Fermi velocity and the superconducting gap depend on the band index and *k*-vector. This was soon pointed out by *ab initio* calculations [20, 22]. No superconductor up to now has shown so clearly the influence of both band and gap anisotropy on the specific heat in the superconducting state, allowing them to be characterized in the bulk. This is the central point of this article.

This paper is organized as follows. In Section II, as a reminder on specific heat, we briefly recall the overall features of the total specific heat of MgB$_2$ in various temperature ranges, and point out to the information contained in these ranges. In Section III, we describe briefly the special requirements of calorimetric techniques for such studies. In Section IV, the electronic specific heat in zero field is shown to reflect a bimodal distribution of gaps. In Section V, the specific heat of a crystal measured at $T \ll T_c$ as a function of both magnetic field $H$ and orientation is shown to reflect the different mass anisotropy of both groups of bands crossing the Fermi level. In Section VI, we describe the effect of disorder induced by irradiation, and conclude in Section VII.

## II. General features of the specific heat of MgB$_2$

Here we present some general results and briefly emphasize the "anomalies" of MgB$_2$, which will be treated in more details in the following sections. Table I sums up important parameters, compared to those of Nb$_3$Sn, which is one of the A15 superconductors with the highest $T_c$. Figure 1 shows the total specific heat (electrons and phonons) of a ceramic sample of MgB$_2$ from about 1 to 45 K. To a good approximation, the normal-state specific heat $C_n$ is given by the curve measured in a magnetic field of the order of 14 T [23]. The maximum upper critical field is as high as ~ 18 T, but at such fields only one band contributes, and furthermore over a narrow solid angle, so that in practice the specific heat no longer changes significantly beyond ~ 8 T (See Sections IV and V). The intercept of the normal-state curve $C_n/T$ at $T \to 0$ provides the coefficient $\gamma_n$ of the linear term of the electronic specific heat, whereas the slope $d(C_n/T)/d(T^2)$ determines the initial Debye temperature. The ratio of $\gamma_n$ over the bare DOS as given by band-structure calculations [17-19, 21, 24-27] allows the mass enhancement factor $1+\lambda_{e-ph}$ to be determined. Thus one can directly evaluate the average electron-phonon coupling constant $\lambda_{e-ph} \sim 0.6$ with less than 20% uncertainty. Such a value would be typical for superconductors with a much lower $T_c$ (e.g. V, Ta). The Debye temperature $\theta(T)$ is high, $\sim 900$ K at $T \rightarrow 0$ and at room temperature, and has a minimum near 25 K, showing excess weight in the phonon density of states near $\theta(0)/3$ [9]. This is a common feature among metals, of no particular bearing on the superconducting properties of MgB$_2$.

The zero field specific heat (Fig. 1) shows the bulk transition at $\sim 37$ K; the vanishing $C/T$ at $T \rightarrow 0$ establishes that all carriers are paired. Even before subtracting the lattice specific heat, one notices the abnormal negative curvature at $T \rightarrow 0$, and the relatively small size of the specific heat jump at $T_c$ compared to $\gamma_n$. These features reflect the existence of a multicomponent gap, as discussed in Section IV. The condensation energy $E_c$ is obtained by integration of $C_s(T) - C_n(T)$ from 0 to $T_c$. Although somewhat sample-dependent [3, 9], it remains astonishingly small (see Table I). Magnetization measurements on single crystals lead to similar results [28]. The integration of $[C_s(T) - C_n(T)]/T$ from 0 to $T_c$ gives the entropy jump at the transition, which is found to be zero within experimental uncertainty, as expected for a second-order transition.

The electronic specific heat in the mixed state, $C_{es}(H,T)$, is obtained by subtracting the normal-state specific heat $C_n(T)$ from the measured data $C(H,T)$, and then adding the normal-state electronic specific heat $C_{en} = \gamma_n T$. The effect of the magnetic field is most prominent either near $T_c$ or at $T \rightarrow 0$. The upper inset of Fig. 1 shows the electronic component near $T_c$. The transition is shifted, and at the same time broadened by the field. This is a consequence of the angular ($\vartheta$) distribution of critical fields $H_{c2}(\vartheta,T)$ in a ceramic sample, rather than fluctuations. The onset of the transition and the maximum of the anomaly determine two extreme $H_{c2}(\vartheta,T)$ lines with an anisotropy ratio $\Gamma = H_{c2}(\pi/2,T)/H_{c2}(0,T) \approx 3$ near $T_c$ [6]. More detailed understanding of this region is obtained from studies on single crystals [29]. The lower inset of Fig. 1 shows the non-linear increase of the low temperature specific heat, which is a complicated function of $H$ and $T$. At $T \to 0$, the intercept $\gamma(H)$, which represents the occupancy of excited electronic states in the vicinity of vortex cores, undergoes nearly the same change between 0 and 1 T as between 1 and 14 T. The region near 10 K appears as a crossover at very low fields. These features, departing from the ISB model, will be the central point of Sections IV and V. Note that the rapid rise of the superconducting state entropy in low fields at $T \to 0$ must be compensated in some temperature range, since the entropy above $T_c$ must be independent of the path followed in the ($H$, $T$) plane. It is actually balanced by the rapid and unusual decrease of the amplitude of the specific heat jump versus the magnetic field (upper inset of Fig. 1). Therefore these two features are thermodynamically consistent.

### III. Techniques

Various techniques allow a tradeoff to be chosen between contradictory requirements such as small sample mass and high accuracy. Two extreme cases are illustrated by the continuous adiabatic technique, which requires "large" samples weighing typically 0.1 gram or more, but yields data with 0.1 to 0.5% absolute accuracy [30], and AC calorimetry, which works with milligram to sub-microgram samples, but only gives relative data, to be calibrated at some point [31, 32]. However, a common point is that both techniques may be highly reproducible, and able to resolve small anomalies, the scatter being as low as 0.01%. The relaxation technique appears as a compromise [33, 34].

Ceramics are not limited in size. We measured the specific heat of the ~ 0.1 g sample shown in Fig. 1 at high temperature (20-250 K) using an adiabatic technique, and at low temperature (1-25 K) with a relaxation technique (see Ref. [6, 9, 35] for more details). On the other hand, presently available single crystals weigh at most ~ 200 µg. The 39 µg crystal discussed in Section IV (see Ref. [36] for information on crystal growth) was measured at low temperature (< 3 K) with a specially designed relaxation calorimeter, using signal averaging to reduce the noise [37].

One difficulty to be overcome is caused by the high Debye temperature of MgB$_2$. Accordingly, the specific heat in zero field is very small at low $T$. Therefore, thermally conducting adhesives commonly used in calorimetry add a large and somewhat uncertain contribution to the total heat capacity, which must be subtracted. Depending on the sample/addenda ratio, this subtraction may be the main source of error in final absolute values of $C$, whereas it does not affect relative changes versus $H$. Furthermore, the specific heat may increase by several orders of magnitude with $H$ or $T$. This restricts the useful range of data obtained by a single technique. Results given by fully automated apparatus may have to be considered with caution. Finally, studies of the electronic specific heat require high fields; note that the maximum $H_{c2}$ can reach ~ 30 T (Section VI).

## IV. Specific heat in zero field

Plots of $C(0,T) - C(H,T)$ nearly show the electronic specific heat difference $C_{es}(T) - C_{en}(T)$ if $\mu_0 H$ exceeds ~ 8 T. The latter quantity, properly normalized as $[C_{es}(T) - C_{en}(T)]/\gamma_n T$ versus $T/T_c$, has been tabulated for an ISB superconductor in the weak-coupling limit [38]. Figure 2 shows both the BCS curve and the data measured for a high density MgB$_2$ ceramics prepared from high purity elements in a cubic anvil press [6, 35]. Differences are observed in the amplitude and sharpness of the specific heat jump, and in the low-temperature upturn due to magnetic impurities, but the large excess weight near $T_c/5$ is a particularly robust feature: similar results have been observed on different samples [9] and by other laboratories [3, 10]. This excess is compensated by a missing weight near $T_c$, as required by the thermodynamics of 2$^{nd}$ order transitions, since the area in a $C/T$ versus $T$ plot is an entropy. Accordingly, the amplitude of the $C$ jump is small. Yang *et al.* [10] and Bouquet *et al.* [3] have noted that the exponential decrease of $C$ at very low temperature is representative of a small gap ~ 1 to $1.5 k_B T_c$, whereas Kremer *et al.* [7] have reported that the shape of the transition at $T_c$ suggests a gap ~ $4 k_B T_c$ and strong coupling.

These unusual and in part apparently contradictory features are a consequence of the particular band structure of MgB$_2$. *Ab initio* calculations soon proposed an explanation for these puzzles by pointing out to the importance of anisotropy. Two main groups of bands cross the Fermi energy. One of them consists of nearly three-dimensional (3D) electrons and holes, represents ~ 56% of the total DOS, and is referred to as the π-band. The other one, the σ-band, consists of nearly 2D holes, and is responsible for ~ 44% of the total DOS. These different carriers give rise to two groups of gaps clustered around the values $\Delta_{0,\pi} \cong 2$ and $\Delta_{0,\sigma} \cong 7$ meV at $T = 0$. Both gaps close at the same $T_c$ [19, 20, 39, 40]. Thus, MgB$_2$ belongs to the family of multi-band superconductors: each band carries a superconducting gap with a different amplitude. This can be considered as a particular case of gap anisotropy.

It was shown how the specific heat of such a system can be modeled and interpreted phenomenologically in a simple way [41]: MgB$_2$ is considered as a mixture of two superconductors, a fraction $\gamma_{\pi,n}/\gamma_n$ being characterized by a normalized gap ratio $2\Delta_{0,\pi}/k_B T_c$, the remaining fraction $\gamma_{\sigma,n}/\gamma_n$, by $2\Delta_{0,\sigma}/k_B T_c$. Technically, the specific heat of such a system is calculated as shown 30 years ago by the so-called α-model [42], assuming that the temperature dependence of each gap does not depart much from the normalized BCS curve. Depending on samples, the results for the three parameters of the fit are in the range $\gamma_{\pi,n} : \gamma_{\sigma,n} \cong 50{:}50$ to $55{:}45$, $2\Delta_{0,\pi}/k_B T_c \cong 1.2$ to $1.3$, and $2\Delta_{0,\sigma}/k_B T_c \cong 3.8$ to $4.4$ [41]. The quality of the fit shows that no additional free parameter is required (Fig. 2). The superfluid density given by measurements of the penetration depth can be successfully analyzed using a similar approach [43].

The smaller gap $\Delta_\pi(T)$ is the main source of the excess specific heat at low $T$. Indeed, let us consider the case of a semiconductor with an electronic gap equal to $\Delta_{0,\pi}$. At low $T$, its specific heat shows the characteristic exponential dependence, but when the temperature approaches and exceeds $\Delta_{0,\pi}/k_B$, $C/T$ saturates (Fig. 3). It is meaningless to consider such thermal excitations with energy larger than the superconducting gap in the ISB model, since the energy scales $k_B T_c$ and $\Delta_0$ overlap: the transition occurs (i.e., the gap closes) when the thermal energy $k_B T$ is of the order of half the zero-temperature gap. For MgB$_2$, on the contrary, both energy scales are well separated for the π-band, and a semiconductor-like situation arises: when the gap closes, $k_B T$ is large compared to $\Delta_{0,\pi}$. Whereas a single parameter $T_c$ is sufficient to describe an ISB superconductor, *two* scales are needed for MgB$_2$: $T_c$, at which the gap closes, characterizing $\Delta_{0,\sigma}$, and $T_{c,\pi}$, characterizing $\Delta_{0,\pi}$, defined as the critical temperature of a virtual ISB superconductor with the same gap value ($T_{c,\pi} = \Delta_{0,\pi}/1.76 k_B \approx 13$ K). Figure 3 shows the contribution of the π-band to $C_{es}$, calculated within the α-model: at the crossover temperature $T_{c,\pi}$, the thermal excitations become large enough to generate the excess of specific heat observed experimentally. At $T = T_c \gg T_{c,\pi}$, the superfluid condensate of the π-band carriers is thermally depleted, so that closing of the π-gap only gives rise to a minute jump.

On the other hand, the specific heat associated with the larger gap $\Delta_{0,\sigma}$ more closely resembles the usual BCS curve. Since only about 50% of $\gamma_n$ is associated with this gap, the jump for the whole system is reduced in the same proportion, so that $\Delta C/\gamma_n T_c < 1.43$. This reduced jump has often been considered as the main characteristic feature to be explained quantitatively by theory. From the experimental point of view, artifacts (e.g. metallurgical inhomogeneity) might also lead to a reduction of the jump, so that the most robust features to be accounted for, are the exponential variation at very low temperature showing an abnormally small gap, and the hump in the specific heat near $T_c/4$.

This bulk confirmation of the two-gap model by specific heat is particularly conclusive in MgB$_2$, owing to the fact that both bands contribute to the total DOS with nearly equal weights, and that both gaps do not differ too much in width. However, remember that two-gap superconductivity has been considered theoretically [44, 45] and observed experimentally [46, 47] long ago. More recently, NbSe$_2$ [48] and borocarbides [49] have been interpreted as multigap superconductors. From the theoretical point of view, much of the observed features can be captured by the weak-coupling BCS theory with proper inclusion of gap anisotropy in the most general sense [50]. In particular, in the two-gap scheme, one gap is predicted to be larger and the other one smaller than $1.76 k_B T_c$ [51], as confirmed in MgB$_2$. Therefore gap ratios $2\Delta_0/k_B T_c$ moderately in excess of 3.5 do not necessarily have any bearing on strong-coupling in this case. In the weak-coupling regime, the fitting variables $x \equiv \gamma_{\pi,n}/\gamma_n$, $2\Delta_{0,\pi}/k_B T_c$, and $2\Delta_{0,\sigma}/k_B T_c$ are not independent, but constrained by [51]:

$$x\left(\frac{\Delta_{0,\pi}}{1.76 k_B T_c}\right)^2 \ln\left(\frac{\Delta_{0,\pi}}{1.76 k_B T_c}\right) + (1-x)\left(\frac{\Delta_{0,\sigma}}{1.76 k_B T_c}\right)^2 \ln\left(\frac{\Delta_{0,\sigma}}{1.76 k_B T_c}\right) = 0 \qquad (1)$$

For a test, one can use the fitted values of $x$ and $2\Delta_{0,\pi}/k_B T_c$ to extract $2\Delta_{0,\sigma}/k_B T_c$ from Eq. (1). This yields the value of the larger gap compatible with weak coupling. It is compared with independent determinations of the same quantity in Table II. *Ab initio* calculations and Geneva data appear to be consistent with the weak-coupling regime, whereas Berkeley data and penetration depth experiments seem to require strong-coupling corrections.

Let us finally mention that other types of anisotropy have been considered [52, 53].

## V. Specific heat in the mixed state

The previous results raise the interesting question of the nature of a vortex in a two-gap system, which features two subsystems in *k*-space with widely different characteristics with respect to dimensionality and pairing energy. Indeed, the mixed-state specific heat of MgB$_2$ presents unusual properties. This section summarizes recent experiments performed in Geneva [37] on a crystal grown in ISTEC, Tokyo [36]. Preliminary work performed on ceramics pointed to the extremely fast increase of $\gamma(H)$ at fields much lower than $H_{c2}$ (Fig. 4d) [3, 6, 9, 10]. In order to compare this behavior with well-documented cases, we present in the other panels of Fig. 4 $\gamma(H)$ for Nb$_{77}$Zr$_{23}$ [54], 2H-NbSe$_2$ [55], and YBa$_2$Cu$_3$O$_7$ [34]. For dirty isotropic type-II superconductors such as the alloy Nb$_{77}$Zr$_{23}$, $\gamma(H)$ is nearly linear in $H$ up to $H_{c2}$. This follows from the picture initially given by Caroli, Matricon and de Gennes (CMG) [56, 57] where the contribution of each vortex core to the normal state volume is proportional to $\xi^2$. It has been suggested that some non-linearity results from vortex-vortex interactions [58]. The non-linearity in the case of NbSe$_2$ and borocarbides is related to anisotropy. Finally, the law $\gamma(H) \propto H^{0.5}$ has been documented for *d*-wave superconductors [34, 59]. For YBa$_2$Cu$_3$O$_7$, this square-root law, supplemented by the absence of any exponential regime at low $T$ in $C_{es}(H=0,T)$, is due to the presence of lines on the Fermi surface where the gap vanishes [60]; other power laws could be expected for a different topology of the nodes (e.g. in heavy fermions and Sr$_2$RuO$_4$). However, in none of these anisotropic cases does the deviation from linearity approach the extreme behavior of MgB$_2$. The latter does not originate from non-conventional pairing: the presence of nodes can be excluded due to the presence of an exponential regime in $C_{es}(H=0,T)$ at sufficiently low temperature [3, 10].

Figure 5 presents the low-temperature, low-field behavior of *C*/*T* for MgB$_2$ ceramics. It clearly shows that not only a new temperature scale, but also a new field scale, is present. We have shown in the previous section that the new *T* scale can be obtained by considering $T_c$ of a virtual ISB superconductor having a gap $\Delta_{0,\pi}$ at $T \rightarrow 0$. Similarly, we now understand that three characteristic fields can be defined for MgB$_2$: the anisotropic upper critical field with its minimum $H_{c2,\sigma,c}$ and maximum $H_{c2,\sigma,ab}$, plus a third, smaller, crossover field, which can be described qualitatively as $H_{c2}$ of the aforementioned virtual ISB superconductor. Orders of magnitudes for these fields can be derived using textbook formulas (neglecting the distinction between the Ginzburg-Landau and Pippard coherence lengths, and numerical factors of order one):

$$H_{c2,\sigma,ab}(0) \propto \frac{\Phi_0 \Delta_{0,\sigma}^2}{\hbar^2 v_{F,\sigma}^2} \left(\frac{m_c^*}{m_{ab}^*}\right)^{1/2}$$

$$H_{c2,\sigma,c}(0) \propto \frac{\Phi_0 \Delta_{0,\sigma}^2}{\hbar^2 v_{F,\sigma}^2} \qquad (2)$$

$$H_{c2,\pi}(0) \propto \frac{\Phi_0 \Delta_{0,\pi}^2}{\hbar^2 v_{F,\pi}^2}$$

$H_{c2,\sigma,ab}$ and $H_{c2,\sigma,c}$ are determined by the large gap associated with the σ-band. Its anisotropy can be described by an effective mass tensor with principal axes $m_c^*$ and $m_{ab}^*$; $v_{F,\sigma}$ must then be understood as the Fermi velocity in the (ab)-plane on the Fermi surface sheets shaped as slightly distorted cylinders along the $c$ direction. The ratio $H_{c2,\sigma,ab}(0)/H_{c2,\sigma,c}(0)$ is just the square root of the mass anisotropy factor $\Gamma = (m_c^*/m_{ab}^*)^{1/2} \cong 6$ [61, 62]. $H_{c2,\pi}(0)$ is determined by the small gap associated with the π-band, with nearly isotropic $v_{F,\pi}$. The Fermi velocities for the σ- and π-bands are similar and fall in the range 4.4 to 6.2×10$^5$ m/s [62], so that the ratio $H_{c2,\sigma,c}(0)/H_{c2,\pi}(0)$ is essentially given by the ratio of the gaps $(\Delta_{0,\sigma}/\Delta_{0,\pi})^2 \cong 10$. It follows that starting from $\mu_0 H_{c2,\sigma,ab}(0) \cong 18$ T, we expect $\mu_0 H_{c2,\sigma,c}(0) \cong 3$ T, and $\mu_0 H_{c2,\pi}(0) \cong 0.3$ T in a first approximation.

In order to describe qualitatively the expected behavior of γ($H$), let us first consider the contributions $\gamma_\pi(H)$ and $\gamma_\sigma(H)$ of both bands as independent. Within the simplest CMG model, both contributions should be linear in field, up to their respective normal-state values $\gamma_{\pi,n}$ and $\gamma_{\sigma,n}$ (with $\gamma_{\pi,n} \sim \gamma_{\sigma,n} \sim \gamma_n/2$, see Section IV). Both contributions saturate at their respective upper critical fields, $H_{c2,\pi}$ and $H_{c2,\sigma}$ (the latter depending on the field orientation). Therefore, the full γ($H$) curve should present two breakpoints before saturating at $\gamma_n$. Furthermore, since at low field (e.g. 0.1 T) and low temperature (e.g. $T_c/20$) the contribution of the π-band is ~10 times larger than that of the σ-band, it would mask the anisotropy of the latter. Therefore, γ($H$) should be isotropic at low $H$. On the contrary, beyond $H_{c2,\pi}(0)$, when $\gamma_\pi(H)$ saturates, the anisotropy of the σ-band should progressively dominate the behavior of γ($H$).

The main approximation in this picture is the independence of the bands. In reality, superconductivity in the π-band is maintained above $H_{c2,\pi}(0)$ by coupling to the σ-band, making $H_{c2,\pi}(0)$ a crossover rather than a critical field. Nevertheless, our measurements on a single crystal of MgB$_2$ do follow the qualitative picture sketched above. Even for polycrystals, in spite of the distribution of grain orientation, one can recognize both slope changes at ~ 0.3 and ~ 3 T (Fig. 4d). Data taken on a single crystal are much more informative (Fig. 6). For $H // c$, $\gamma(H)$ is seen to saturate at $\mu_0 H_{c2,\sigma,c}(0) \cong 3.5$ T, and at $\mu_0 H_{c2,\sigma,ab}(0) \cong 19$ T for $H // (ab)$. A short extrapolation is needed in the latter case, owing to the limited field of the experiment; but the saturation value $\gamma_n$ is already known from the data for $H // c$, and agrees with independent determinations on ceramics. The linear part just below $H_{c2,\sigma,c}$ or $H_{c2,\sigma,ab}$ extrapolates to $\gamma_{\pi,n} \cong \gamma_n/2$ for $H \to 0$, showing that the contribution characterized by the large critical fields is associated with about half the DOS. An enlargement of the low-field data (Fig. 6, inset) shows that the contribution characterized by the smallest critical field is nearly isotropic. The crossover field $\mu_0 H_{c2,\pi}(0) \cong 0.3$-$0.4$ T can be defined by the construction shown in the inset of Fig. 6, i.e. $H_{c2,\pi}(0)$ is the field at which the extrapolation of the initial linear increase of $\gamma(H)$ meets the value $\gamma_{\pi,n} \cong \gamma_n/2$.

The main unknown is the exact shape of $\gamma_\pi(H)$ curve near and above the crossover field $H_{c2,\pi}(0)$. To better define it, we have subtracted the idealized CMG behavior of the σ-band from the total $\gamma(H)$ curve. This can be done for both field orientations with respect to the *c*-axis. The remainder is the mixed-state specific heat of the π-band. It displays a smooth crossover from a CMG behavior at low fields to a saturation at its normal-state value. Within experimental error, it remains isotropic at all fields. The picture that emerges is qualitatively consistent with the numerical calculations of the local DOS (LDOS) in the mixed state performed by Nakai and coworkers [63]. Superconductivity in the π-band manifests itself by vortex cores with a large diameter, $\xi_\pi = \hbar v_{F,\pi}/\pi \Delta_{0,\pi} \approx 50$ nm, which start overlapping at $H_{c2,\pi}$. At higher field, they do not vanish, but the π-band LDOS between vortices rises toward nearly its normal-state value, causing the saturation of $\gamma_\pi(H)$ for $H > H_{c2,\pi}$. This picture was recently confirmed by scanning tunneling spectroscopy [64]. The analysis of the low-*T* thermal conductivity versus magnetic field is also consistent with this picture [65]. The thermal conductivity of electronic origin, being proportional to $C_e v_F^2 \tau$, reflects the behavior of the specific heat, within uncertainties associated with the relaxation time $\tau$ [66].

In summary, $C(H, \vartheta, T = 0)$ experiments on a single crystal have determined $H_{c2,\sigma,ab}(0)$, $H_{c2,\sigma,c}(0)$, $H_{c2,\pi}(0)$, $\gamma_{\pi,n}$, and $\gamma_{\sigma,n}$. Independently, isotropic $C(H = 0, T)$ experiments have determined $\Delta_{0,\pi}$, $\Delta_{0,\sigma}$, and $\gamma_{\pi,n}/\gamma_{\sigma,n}$. Consistency has been found between both sets, with $\gamma_{\pi,n}/\gamma_{\sigma,n}$ close to one, and $\Delta_{0,\sigma}/\Delta_{0,\pi} \cong [H_{c2,\sigma,c}(0)/H_{c2,\pi}(0)]^{1/2} \cong 3$. Finally, the ratio $H_{c2,\sigma,ab}(0)/H_{c2,\sigma,c}(0) \cong 6$ is consistent with various independent determinations of the upper critical fields in the low temperature range [28, 29, 67-69]. The presence of 3D and 2D bands, with different associated gaps, makes it impossible to define a single anisotropy factor for MgB$_2$. Only an effective anisotropy $\Gamma(H, T)$ can be defined, which depends on the physical quantity that is used to determine it. In our case, $\gamma(H, \vartheta)$ is essentially a measure of the area occupied by vortex cores, as determined by $\xi_\pi$, $\xi_{\sigma,ab}$, and $\xi_{\sigma,c}$. Our measurements are consistent with an isotropic coherence length in the $\pi$-band, and a coherence length governed by the anisotropy of the effective masses in the $\sigma$-band. Note that this anisotropy needs not be the same as that of the penetration depth [61, 70, 71].

The degree of consistency between various experiments on one hand, and between experiment and theory on the other hand, provides substantial support to the two-band picture for MgB$_2$. The understanding of the mixed state of MgB$_2$ suggests to revisit other anisotropic s-wave systems for which an unusual behavior of $C_s(H, T)$ has been reported.

## VI. Effect of disorder induced by irradiation

One of the predictions based on the band structure of MgB$_2$ refers to the particular influence of impurities on $T_c$. Whereas intraband scattering does not change the two gaps (Anderson's theorem), interband scattering does. A large amount of impurity scattering will cause the two gaps to converge to the same value [72], estimated as $\Delta_{0,\sigma} = \Delta_{0,\pi} = 4.1$ meV, corresponding to $T_c \cong 25$ K and $2\Delta_0/k_B T_c \cong 3.7$ [62].

There is no easy experimental control of the inter- versus intraband scattering rate. The relative insensitivity of $T_c$ to sample quality, from commercial powders to high purity single crystals, seems to show that intraband scattering is commonly acting [73]. However, stronger disordering by irradiation can lead to a drastic reduction of $T_c$ [74], suggesting interband scattering. This motivated our study where the behavior of both gaps is monitored by bulk specific-heat experiments while disorder is introduced in the material. With a minimum $T_c$ of 30 K, the point of convergence of the gaps was not reached, so that this part of the work remains tentative. However, significant quantitative changes are observed. Details are given in Ref. [75].

The polycrystalline sample for which results have already been shown in Fig. 1 was irradiated at the Triga reactor in Vienna in two steps, first to a fast neutron fluence (E > 0.1 MeV) of $1\times10^{22}$ m$^{-2}$, later completed up to a total of $3\times10^{22}$ m$^{-2}$. The introduced disorder is estimated to 0.05 displacement per atom in the final state. Thermal neutrons, which would damage only the surface of the sample, were shielded out by cadmium in order to obtain an homogeneous defect structure.

The specific heat jump remains sufficiently well-defined upon irradiation (Fig. 7). Note that specific heat determinations of $T_c$ set the highest requirements on sample homogeneity, reflecting the true superconducting volume, whereas Meissner effect, magnetic shielding and resistivity may show full transitions caused by a minority fraction of the sample. This is of particular importance in MgB$_2$ for which surface superconductivity has been reported [76]. In zero field, the bulk critical temperatures were 37.0 K before irradiation, 34.1 K at an intermediate stage, and 30.2 K after irradiation. In a magnetic field, broadening occurs essentially due to anisotropy. Transition onsets, which mark the $H_{c2,\sigma,ab}(T)$ line of the phase diagram, remain well defined as long as $T_c$ is reduced by no more than ~ 10 K, i.e. in fields up to 3-8 T (Fig. 1 and 8, insets). In this interval, a significant increase of the average slope $\left(dH_{c2,\sigma,ab}(T)/dT\right)_{T_c}$ is observed with irradiation. These determinations of $T_c$ were continued by determinations based on magnetotransport up to 28 T in a Bitter-type magnet at the GHMFL, Grenoble. The midpoint of the resistance steps occurs slightly below the onset of the specific heat jumps in the overlap region. The estimated values of $\mu_0 H_{c2,\sigma,ab}(0)$ are 18 T in the pristine state and slightly above 28 T after the second irradiation (Fig. 8). We have verified that $\gamma_n$ remains nearly constant during the process, in contrast to the residual resistivity just above $T_c$, which increases by a factor of ~ 6 (4.0 and 23 $\mu\Omega\cdot$cm, respectively). This suggests that the main mechanism enhancing $H_{c2,\sigma,ab}(0)$ is the reduction of the mean free path. Further analysis is complicated by the fact that in a two-gap situation, π-band carriers may be nearly in the dirty limit whereas σ-band carriers may be in the clean limit, owing to their different coherence lengths.

The specific heat in the superconducting state was analyzed in terms of the two-gap model as in Section IV. Upon disordering, the smaller gap $\Delta_{0,\pi}$ is found to remain almost unchanged, increasing from 2.1 to 2.2 meV. The larger gap $\Delta_{0,\sigma}$ decreases from 6.2 to 4.7 meV. The relative weight associated with each contribution, $\gamma_{\pi,n} : \gamma_{\sigma,n}$, remains between 50:50 and 55:45. In terms of reduced gap values, a tendency to convergence is observed since $2\Delta_{0,\pi}/k_B T_c$ increases from 1.3 to 1.7 and $2\Delta_{0,\sigma}/k_B T_c$ decreases from 3.9 to 3.6, but both sets of gap values still remain well separated, so that the isotropic meeting point seems to occur for $T_c$ much lower than 30 K (Fig. 8, inset). The absence of any significant variation of the reduced specific heat jump $\Delta C/\gamma_n T_c \cong 0.90$ to 0.85 is another indication that convergence of both gaps is still far from being reached, since the ISB value $\Delta C/\gamma_n T_c = 1.43$ would be expected at the meeting point. In conclusion, the two-gap features are more robust than expected from the theory of interband scattering [62, 77].

Although this work was performed on ceramics, we can state that irradiation-induced scattering increases not only $H_{c2,\sigma,ab}$, but also $H_{c2,\pi}$. This is demonstrated by a plot of the specific heat $C/T$ at $T \ll T_c$ versus $H$ (Fig. 9). As discussed in the Section V, the contribution at low field ($< 0.5$ T) and low temperature mostly comes from the isotropic π-band, $\gamma_\pi(H)$. Owing to the polycrystalline nature of the sample, the breakpoints at the three crossover / critical fields $H_{c2,\pi}(0)$, $H_{c2,\sigma,c}(0)$, and $H_{c2,\sigma,ab}(0)$ are smeared into a smooth, quasi-logarithmic $H$-dependence (Fig. 9). The remarkable point is that the whole curve is shifted almost parallel to itself, even at low fields, the field scale being increased by a factor of nearly two. This means that not only $H_{c2,\sigma,ab}(0)$, but also $H_{c2,\pi}(0)$ is nearly doubled. This property might be understood within the scenario suggested by STS, in which π-band superconductivity does not exist independently, but is induced by σ-band superconductivity [64].

## VI. Conclusion

The specific heat of MgB$_2$ has shown unconventional thermodynamics in the superconducting state. The contribution of the smaller gap is reminiscent of that of a semiconductor, and gives rise to a large excess specific heat in the vicinity of $T_{c,\pi} \ll T_c$. The mixed-state specific heat is also unconventional, showing unprecedented non-linearity of $C/T$ versus $H$ at low temperature; e.g. with a field of only 5% of $H_{c2,\sigma,ab}(0)$, nearly 50% of the normal-state value $\gamma_n$ is restored. This peculiar behavior is described phenomenologically by three characteristic fields, one isotropic crossover field $H_{c2,\pi}$ associated with the 3D $\pi$-band, and one anisotropic critical field with extreme values $H_{c2,\sigma,c}$ and $H_{c2,\sigma,ab}$ associated with the quasi-2D $\sigma$-band.

The thermodynamics of a multiple-gap superconducting system has already been considered theoretically 40 years ago (see Ref. [45] and references therein), but never before was such a clear example available for experimental study. Isotropic averages such as the mass renormalization constant $m^*/m = 1 + \lambda_{e-ph}$, the total DOS at the Fermi level, or the Debye temperature, are unable to describe the superconducting and thermodynamic properties of MgB$_2$. Instead, one must distinguish band contributions to the electronic specific heat, upper critical and crossover fields $H_{c2,\sigma,c}$, $H_{c2,\sigma,ab}$, and $H_{c2,\pi}$, intra- and interband coupling constants and scattering rates, etc. With this in mind, a high degree of consistency between predictions of *ab initio* numerical calculations and bulk experiments can be obtained.

With regard to applications, technologically useful properties of MgB$_2$ such as the high critical temperature, the large critical current density, and the relatively simple metallurgy, have been partly offset by the limited upper critical field initially reported to be ~ 15 T, much below the Clogston paramagnetic limit which lies near 70 T. The remarkable bulk increase of $H_{c2}$ obtained by irradiation, as well as by other types of disorder [78, 79], shows that this is not a definitive limit.

## Acknowledgements

The authors thank C. Berthod, A. Carrington, T. Dahm, M. R. Eskildsen, R. A. Fisher, R. Flükiger, A. Holmes, J. Karpinski, J. Kortus, C. Marcenat, M. Mishonov, N. E. Phillips, and E. Schachinger for fruitful discussions, and J. Hinderer, B. Revaz, A. Lo, and B. Seeber for assistance in experiments. This work was supported by the Swiss National Science Foundation through the National Centre of Competence in Research "Materials with Novel Electronic Properties – MaNEP" and the New Energy and Industrial Technology Development Organization (NEDO, Japan).

## Figure captions

Figure 1, main panel: total specific heat $C/T$ versus temperature squared for a ceramic sample of MgB$_2$, in the normal ($\mu_0 H = 14$ and 16 T) and superconducting state ($H = 0$). Upper inset: specific heat difference $[C_s(H) - C_n]/T$ near $T_c$ for fields of 0, 0.5, 1, 2, and 3 T. Lower inset: total specific heat $C/T$ below 16 K for fields of 0, 0.1, 0.3, 1, 2, 4, 8, and 14 T (the latter two data sets are nearly undistinguishable) [6].

Figure 2: electronic part of the specific heat in the superconducting state $C_{es}/T$ normalized to $\gamma_n$, versus reduced temperature. The hatched area marks the low temperature excess with respect to the isotropic single band BCS (ISB) model (thin line). The thick line through the data is the two-gap fit [41].

Figure 3: specific heat in normalized units $C/\gamma_n T$ versus $T/T_c$, first in a simplified model where the smaller gap $\Delta_\pi = \Delta_{0,BCS}/3$ is constant and semiconductor-like (thin line), then in the α-model with $\Delta_\pi(T) = \Delta_{BCS}(T)/3$ (dotted line), compared to the standard ISB behavior (thick line). $T_c = \Delta_{0,BCS}/1.76 k_B$ in all cases.

Figure 4: mixed-state specific heat $C/T$ at a fixed temperature $T \ll T_c$ as a function of the magnetic field. The lines through the data are guide to the eyes. (a), Nb$_{77}$Zr$_{23}$ (extrapolation to $T \to 0$, $T_c = 10.8$ K, $\mu_0 H_{c2} = 7.9$ T) [54]; (b), 2H-NbSe$_2$ ($T = 1.5$ K, $T_c = 7.1$ K, $\mu_0 H_{c2,ab} = 12.5$ T) [55]; (c), YBa$_2$Cu$_3$O$_7$ ($T = 2.2$ K, $T_c = 88$ K, $\mu_0 H_{c2} >$ 100 T) [34]; (d), MgB$_2$ ceramics ($T = 3$ K, $T_c = 38$ K, $\mu_0 H_{c2,ab} = 18$ T) [6].

Figure 5: specific heat $C/T$ as a function of field and temperature for a ceramic sample of MgB$_2$ synthesized under high pressure [35]. The horizontal grid at 0.83 mJ/K$^2$gat

represents the normal-state value $\gamma_n$. The red curve in the $H = 0$ plane represents the ISB curve. The red straight lines in the $T = 0$ plane represent the ideal CMG behavior for $H // $ c and $H // $ (ab).

Figure 6: coefficient of the electronic linear term versus magnetic field applied parallel (□) or perpendicular (●) to the boron planes (single crystal data from Ref. [37]). The long-dashed line represents the normal-state contribution. The short-dashed lines are guides for the eyes. Inset: expanded view of the low-field region: here the long-dashed line represents the partial normal-state contribution of the π-band. The dotted lines represent the estimated contribution of the σ-band alone for both orientations.

Figure 7. Electronic specific heat $C_{es}/T$ versus $T$ with two-gap fits, top: before irradiation, bottom: after irradiation. The former curve is shifted for clarity. The dashed line represents the ISB model with the same $T_c$ and $\gamma_n$ as the sample after irradiation. Inset: variation of both gaps plotted versus $T_c$ (note that the scale shows increasing scattering from left to right). The star (★) is the theoretically predicted convergence point at 25 K [62].

Figure 8. Phase diagram with $H_{c2}(T)$ lines. Onset of the specific heat jump before (■) and after (●) irradiation, midpoint of the resistance drop before (□) and after (○) irradiation. The inset shows the electronic specific heat difference $[C_s(H) - C_n]/T$ versus $T$ near the superconducting transition after irradiation in fields of 0, 0.5, 1, 2, 3, 4, 5, 6, and 8 T (see inset of Fig. 1 for similar data before irradiation) [75].

Figure 9. Total specific heat $C/T$ at $T << T_c$ versus the magnetic field on a logarithmic scale before (○) and after (◆) irradiation [75].

## Tables

Table I: comparison of characteristic parameters (rounded values) for $Nb_3Sn$ and $MgB_2$ (after Ref. [6]). See text for definitions.

|  | $Nb_3Sn$ | $MgB_2$ |
|---|---|---|
| $T_c$ (K) | 18 | 38 |
| $\theta_D(0)$ (K) | 230 | 900 |
| $\gamma_n$ (mJ/K²gat) | 13 | 0.8 |
| $N_{bs}(E_F)$ (eV·at·sp)$^{-1}$ | 1.0 | 0.12 |
| $m^*/m = 1 + \lambda_{e-ph}$ | 2.8 | 1.6 |
| $\mu_0 H_c(0)$ (T) | 0.5 | 0.3 |
| $E_c$ (mJ/cm³) | 110 | 30 |
| $H_{c2}(0)$ (T) | 25 | 18 |

Table II: relative weights $x \equiv \gamma_{\pi,n}/\gamma_n$ and gap ratios for various bulk determinations of the parameters of the two-band model for $MgB_2$. The fits to experimental data are taken from Ref. [41]; the data source is given in the first column. The gap ratio in the fourth column is recalculated from the first two columns according to Eq. (1) and corresponds to the weak-coupling limit.

|  | $x$ | $\dfrac{2\Delta_{\pi,0}}{k_B T_c}$ | $\dfrac{2\Delta_{\sigma,0}}{k_B T_c}$ | $\dfrac{2\Delta_{\sigma,0}}{k_B T_c}$ |
|---|---|---|---|---|
|  | (fit) | (fit) | (fit) | (Eq. (1)) |
| Specific heat [3] | 0.55 | 1.2 | 4.4 | 4.0 |
| Specific heat [9] | 0.5 | 1.3 | 3.8 | 3.9 |
| Specific heat [30] | 0.5 | 1.3 | 3.9 | 3.9 |
| Penetration depth [43] | 0.6 | 1.6 | 4.6 | 4.2 |
| *Ab initio* calculation [20] | 0.53 | 1.3 | 4.0 | 4.0 |

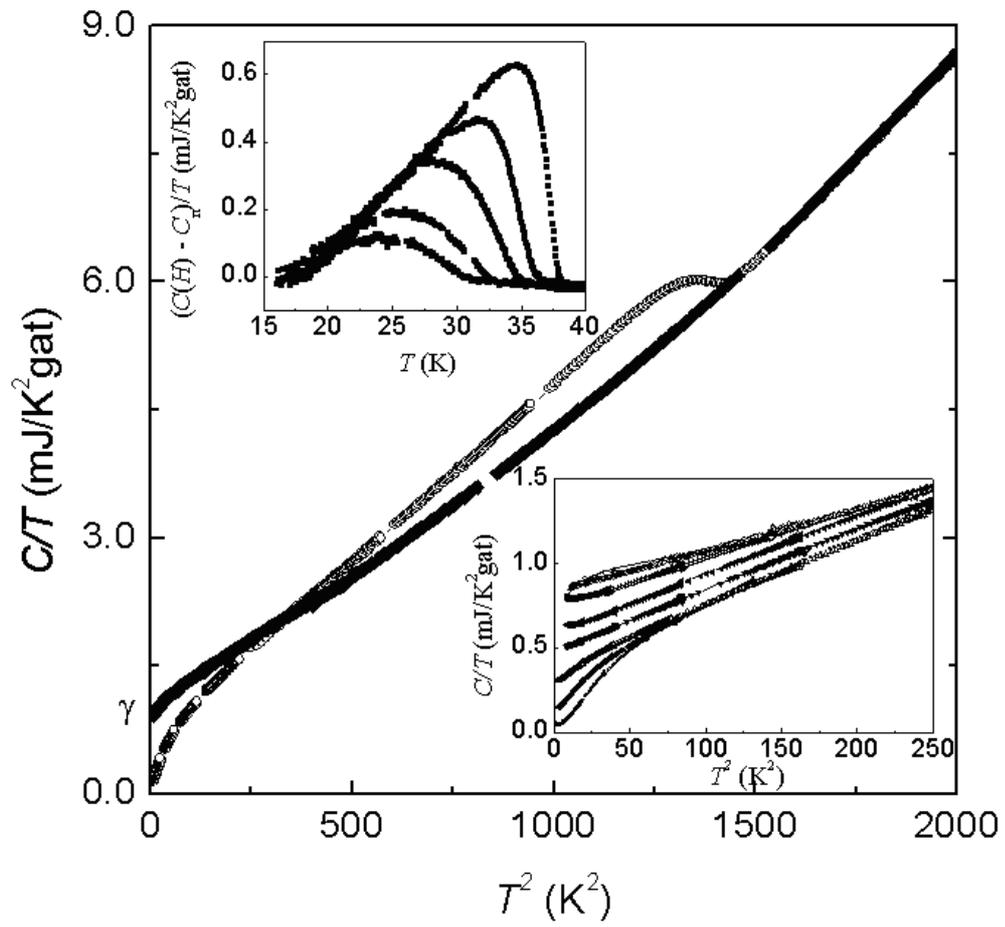

Figure 1

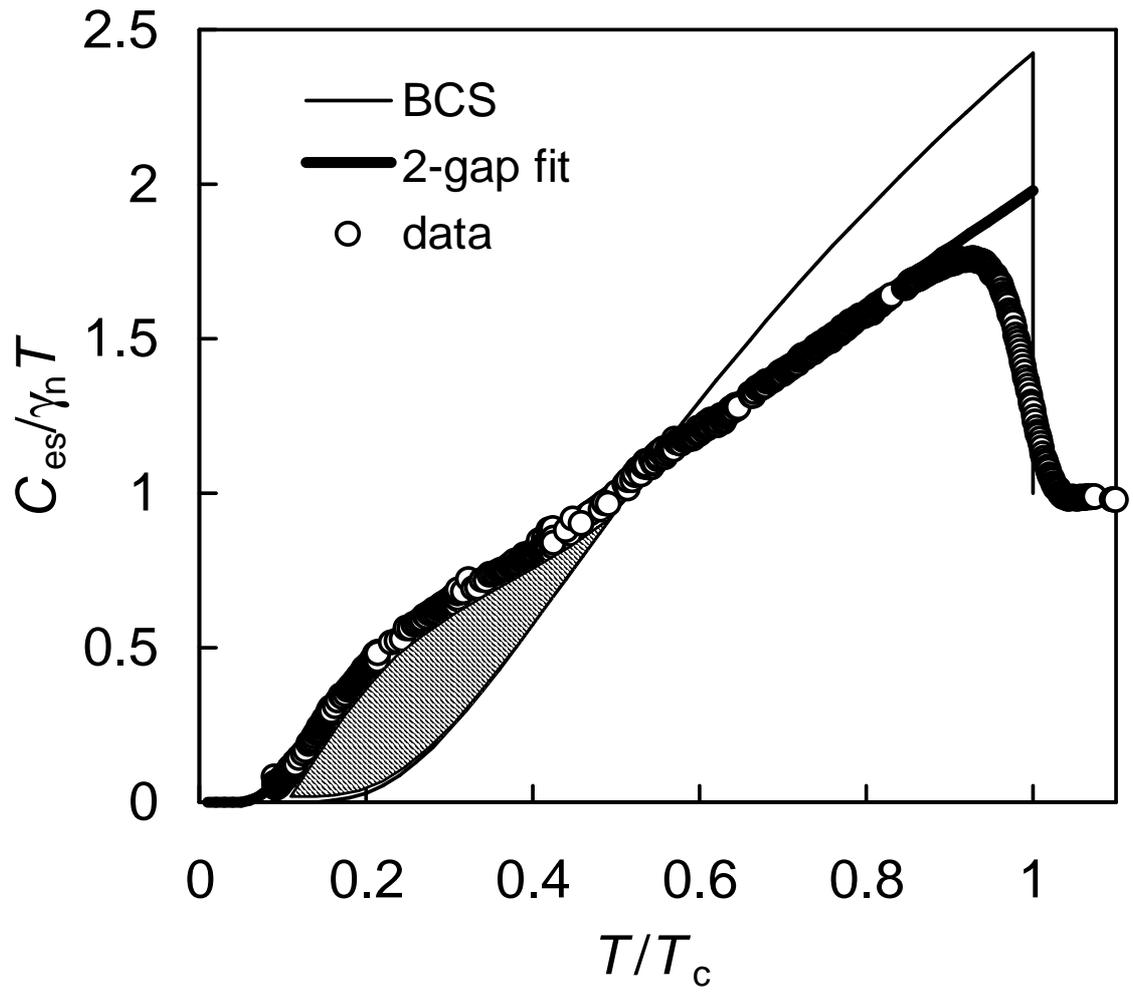

Figure 2

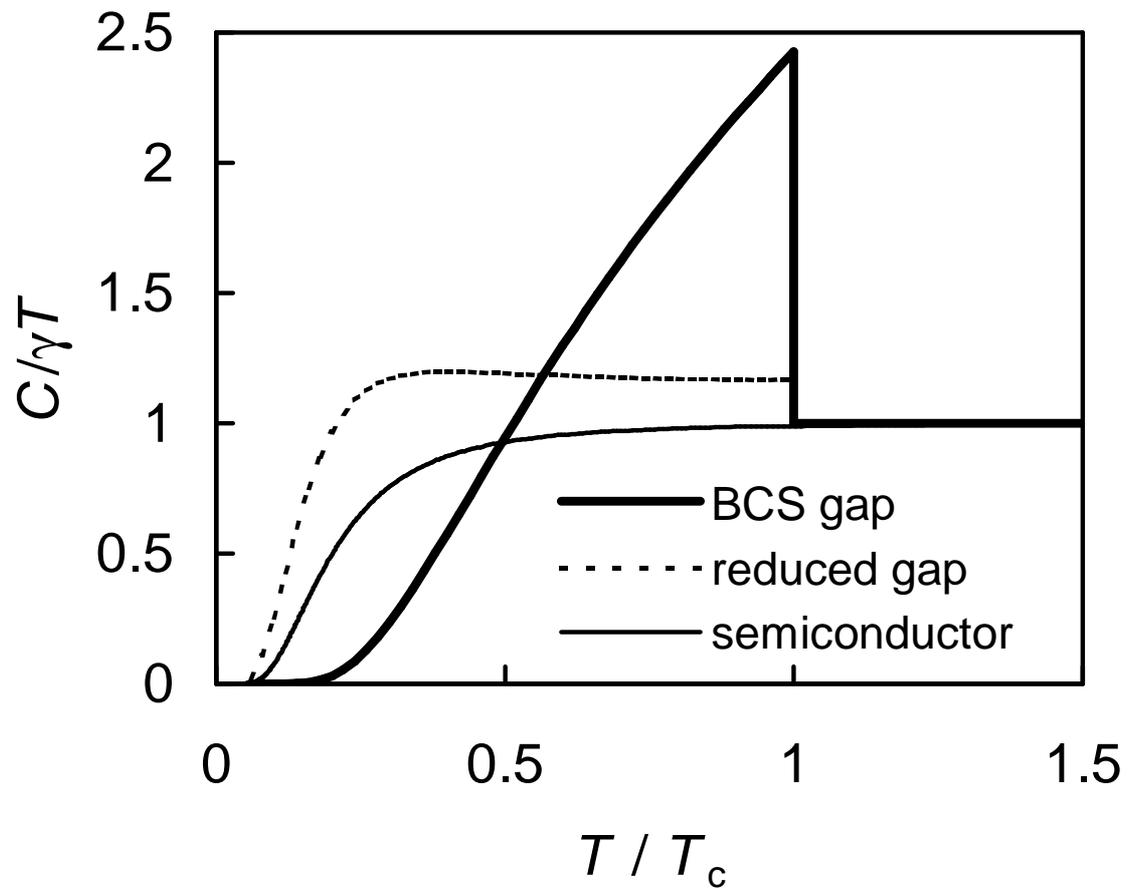

Figure 3

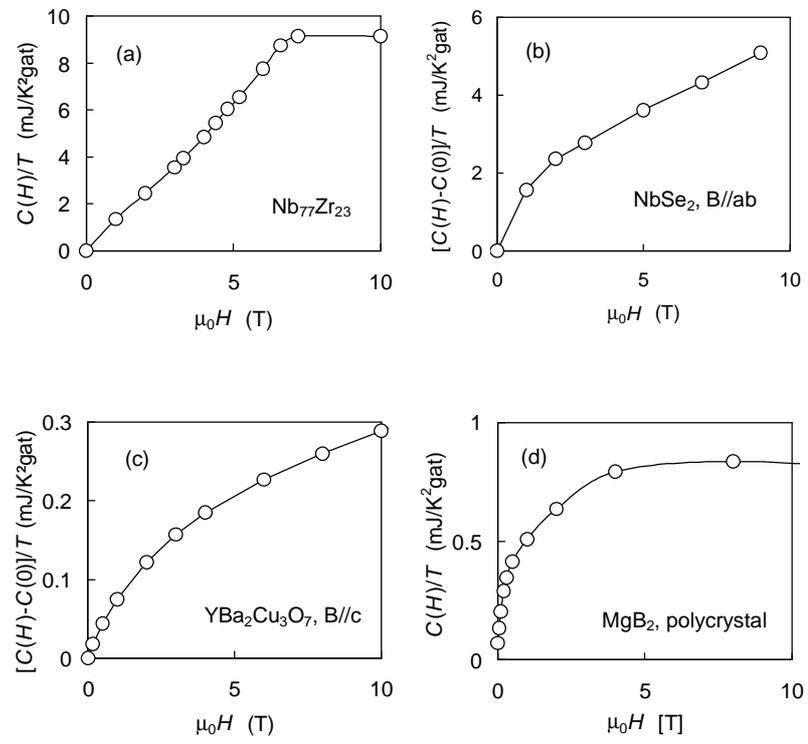

figure 4

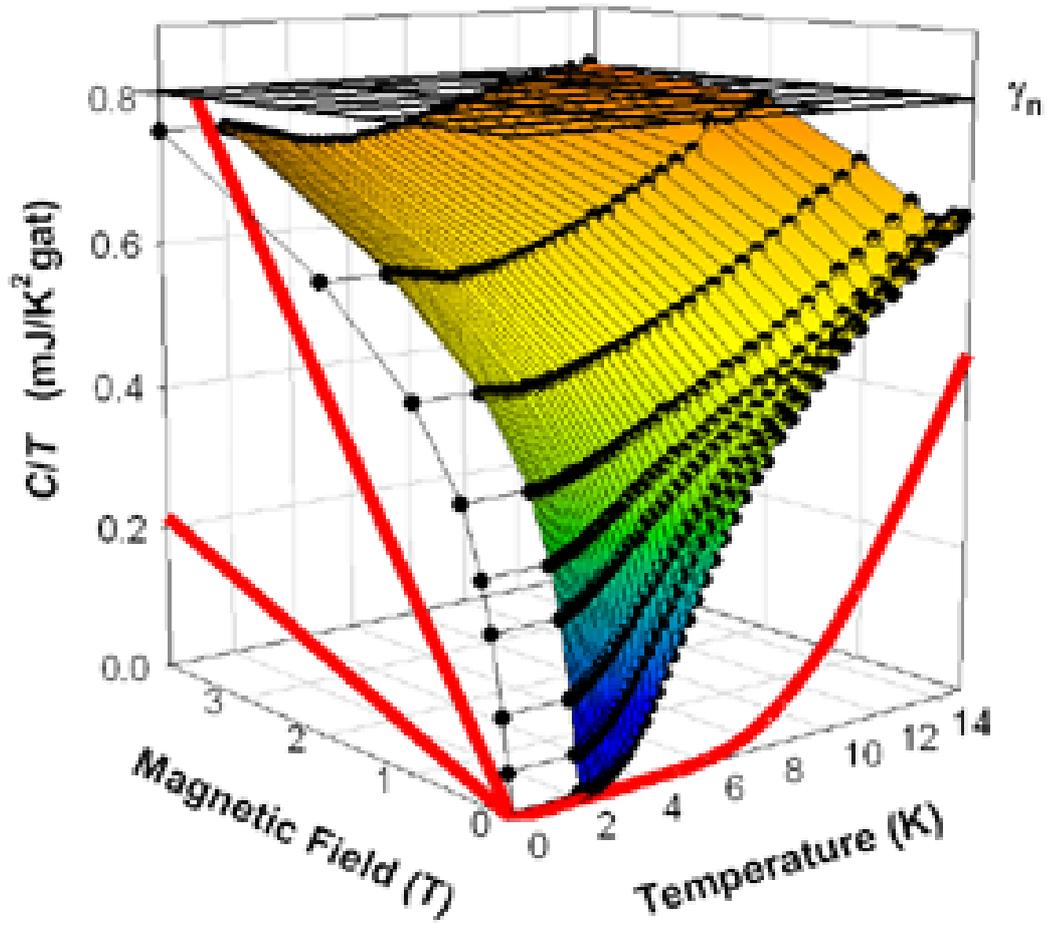

Figure 5

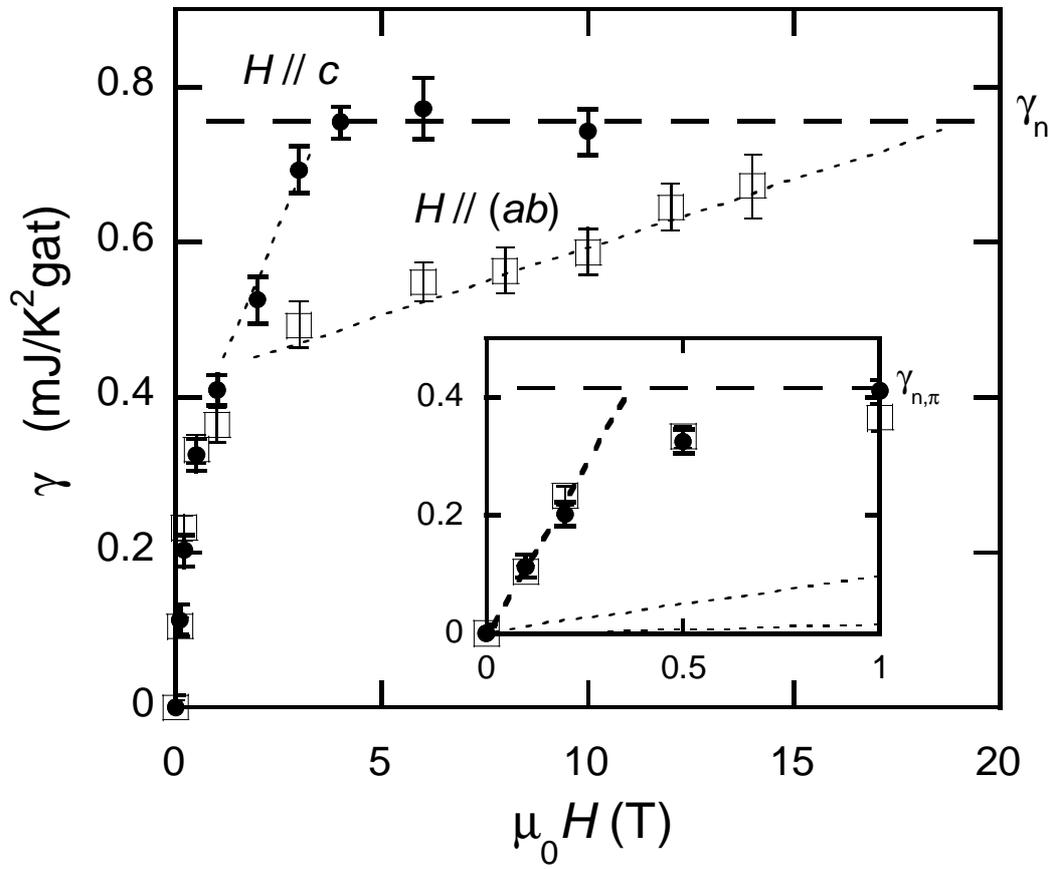

Figure 6

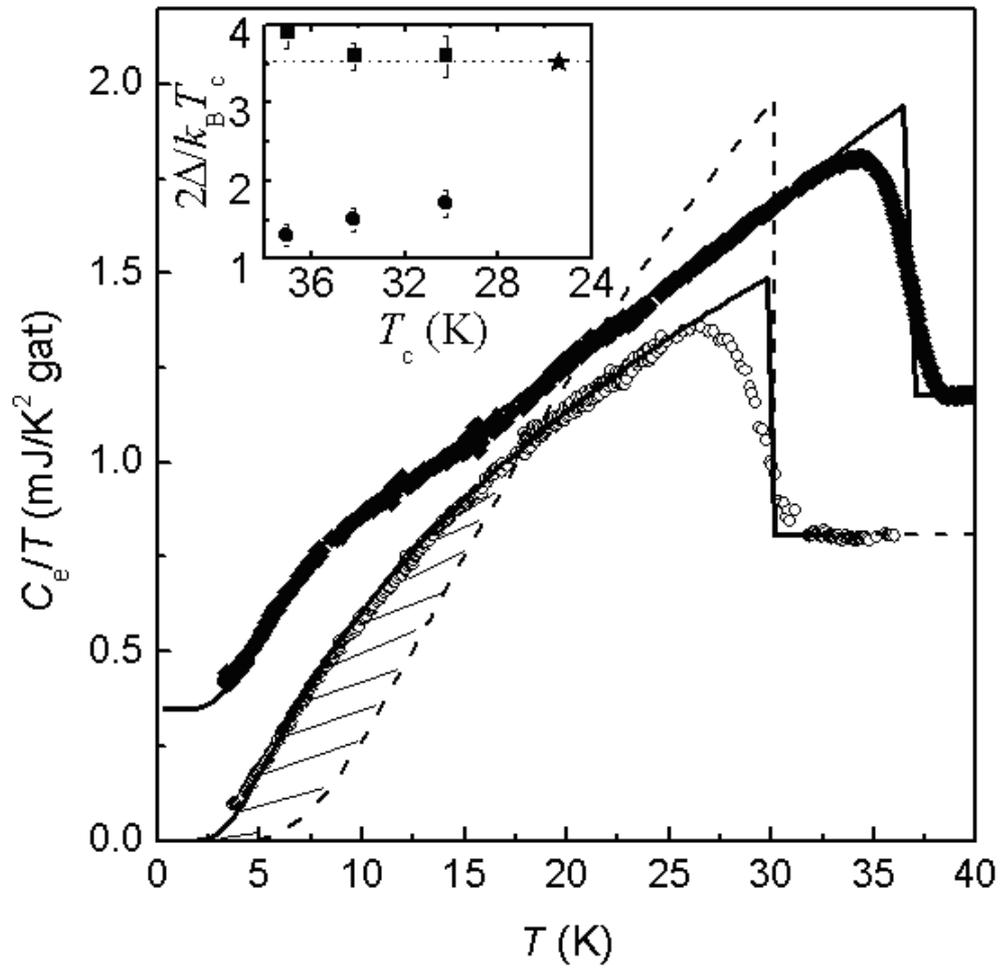

Figure 7

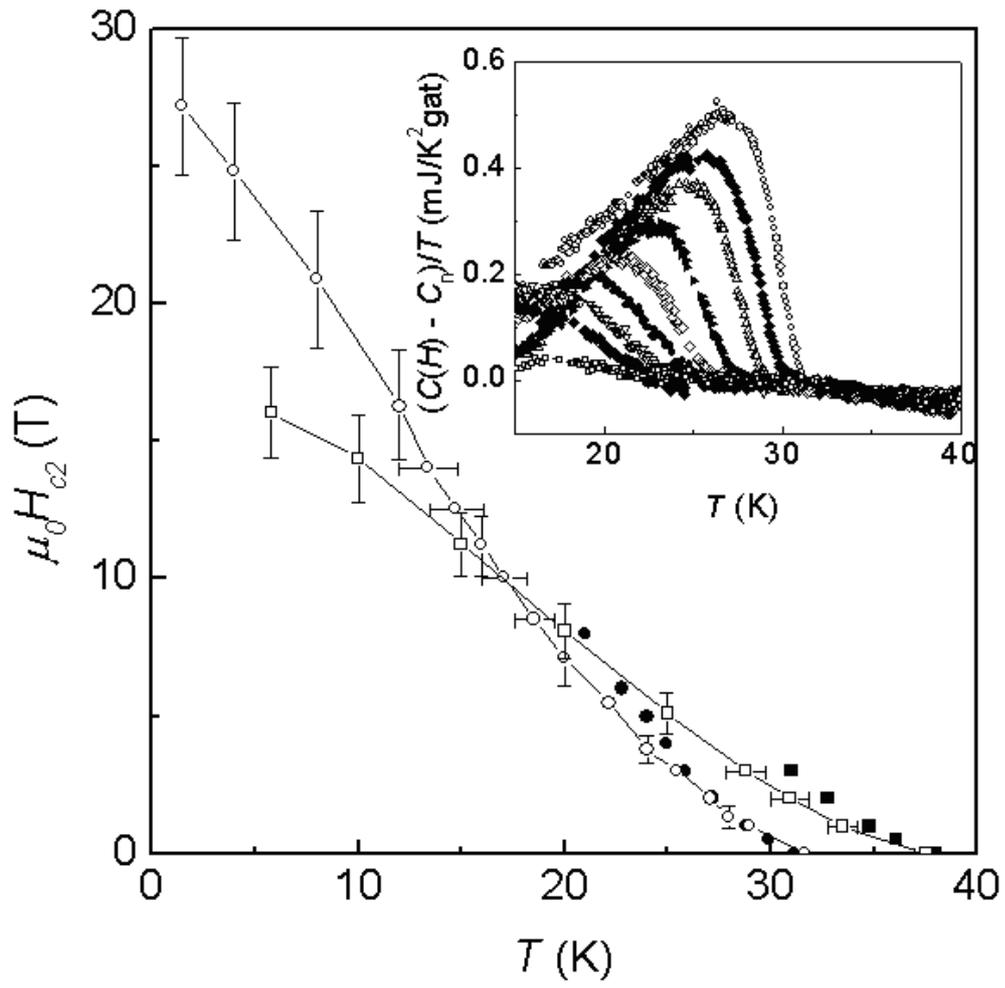

Figure 8

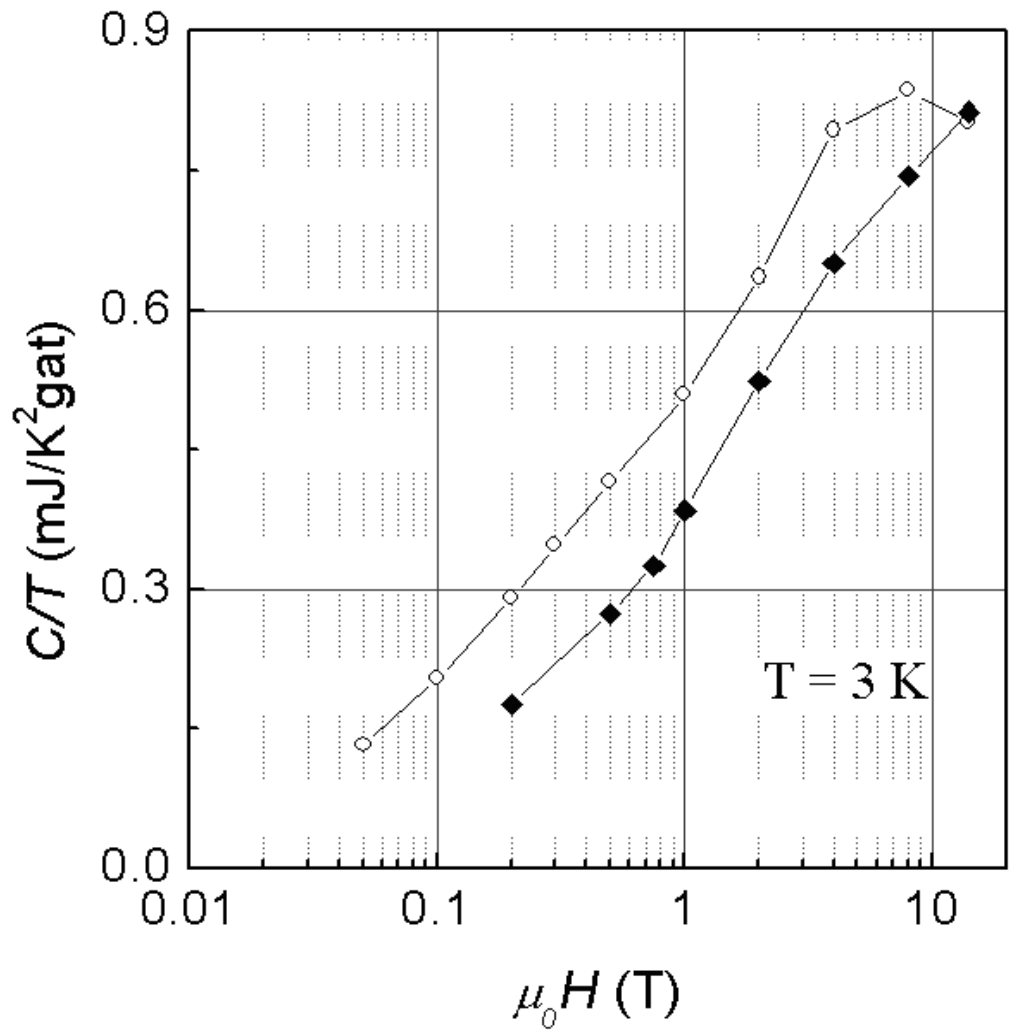

Figure 9